\documentclass[aps,prl,twocolumn,superscriptaddress]{revtex4}
\usepackage{graphics}
\usepackage{graphicx}
\begin{document}
\title{Hydrogen-induced Unzipping of Single-Walled Carbon Nanotubes}
\author{Henry Scudder}
\affiliation{Department of Physics, 
California State University Northridge,
Northridge, CA 91330-8268}
\author{Gang Lu}
\affiliation
{Department of Physics and Division of Engineering and Applied Science, 
Harvard University, Cambridge, MA 02138}
\author{Nicholas Kioussis}
\affiliation{Department of Physics, 
California State University Northridge,
Northridge, CA 91330-8268}
\begin{abstract} 
\vskip 0.5cm
The chemisorption of atomic hydrogen on the single-walled armchair 
carbon nanotube is studied with {\it ab initio} calculations.
A single H atom is found to be chemisorbed on both the inside
and outside wall of the nanotube. The
binding energy of H adsorption at the exterior of the nanotube is 
much greater
than that at the interior of the nanotube.
For the first time, we predict that two rows of H atoms 
chemisorbed on selective sites {\it exterior} to the nanotube can 
break the nearest-neighbor C-C bond of the nanotube
through the concerted formation of C-H bonds, leading 
to the unzipping of the nanotube wall. We provide 
insights into the underlying electronic structure responsible 
for the  H-induced 
unzipping of the nanotube, lending strong support 
to the recent 
experimental observations for the 
coalescence of single-walled nanotubes in the presence of 
atomic hydrogen. Interestingly,
H atoms chemisorbed inside the nanotube do 
not lead to the breaking of the C-C bonds.

\end{abstract}
\maketitle 

Carbon nanotubes have been attracting considerable interest
due to their unique electronic and mechanical properties
since their discovery about a decade ago 
\cite{dresselhaus1}. In particular, there have been intense
studies in
evaluating the capability of carbon nanotubes 
as hydrogen storage material for clean 
energy sources \cite{chen,liu}. It was found that
hydrogen could be stored in bundles of the single-walled nanotubes up 
to 5-10 wt.\% at pressures less than 1 bar near room 
temperature \cite{dillon}. The mechanism, for the most part, 
is attributed to physisorption of H$_2$ molecules both inside 
the tubes and within 
the interstitial regions among the nanotubes at room or
low temperatures \cite{dillon,ye}.
However, recently Liu {\it et al.} observed that, after treating 
carbon nanotubes with H$_2$ gas under high pressure, there was 
residual H$_2$ during the desorption cycle \cite{liu}, suggesting the
presence of chemisorption in the process. On the other hand,
chemisorption 
can also take place during electrochemical storage processes, 
in which H$_2$ molecules are broken into H 
atoms with the application of metal catalysts or electrochemical 
techniques \cite{haufler,becker}. On the theoretical side,
first-principles calculations have confirmed   
the dissociative chemisorption of H$_2$ on two adjacent nanotubes 
in solid phase,
proceeding with the breaking of the 
H-H bond concerted with the formation of 
two C-H bonds on two adjacent 
nanotubes \cite{chan}. The chemisorption of atomic H
on single-walled
nanotubes has also been observed from self-consistent-charge-density 
tight-binding calculations \cite{lee2,lees}. 

Recently, Nikolaev and coworkers discovered
a remarkable phenomenon\cite{smalley}:
under atomic H atmosphere, 
single-walled armchair nanotubes annealed up to 1500$^\circ$C coalesce with 
neighboring tubes, resulting in larger nanotubes with twice and occasionally 
three times the diameter of the original ones. 
Based on these observations, the authors proposed a 
H-activated coalescence mechanism, 
in which the gas-phase H {\it atoms} attack the 
side of neighboring nanotubes, breaking the C-C bonds, 
and producing defective sites on the adjacent nanotubes. 
Once these adjacent defects are formed,  
the strong thermodynamic force, resulting from the released strain 
energy in forming larger tubes, drives the two neighboring smaller tubes to  
join together. Coalescence of clean single-walled nanotubes 
has also been observed under 
electron irradiation conditions at elevated temperatures \cite{terrones}.

In this paper, we present a systematic {\it ab initio} study
of chemisorption of H atoms on
a single-walled (6,6) armchair carbon nanotube.
For the first time, we provide theoretical evidence for 
the H-induced unzipping of the nanotube. Our theoretical
calculations lend strong support to the recent experimental
observations of H-activated coalescence of armchair nanotubes
\cite{smalley}, revealing the electronic mechanism responsible
for this remarkable phenomenon. 

The {\it ab initio} calculations we performed are based on
density-functional theory with  
the CASTEP implementation and ultrasoft pseudopotentials 
\cite{vanderbilt}. 
The Perdew-Burke-Ernzerhof gradient-corrected functional 
\cite{perdew} was used for the 
exchange-correlation potential. 
The energy cutoff for the plane-wave basis was set to 
be 300 eV, yielding a convergence for the total energy 
better than 1 meV/atom.
For the reciprocal-space integration we have used eighteen 
$k$-points along the direction corresponding to the nanotube 
axis using the Monkhorst-Pack scheme
\cite{monkhorst}.
Results have been obtained for the fully relaxed geometries 
including all atoms and the lattice 
constant of the supercell along the tube axis.

The single-walled (6,6) armchair carbon nanotube is modeled 
with a supercell of dimension 
20\AA $\times$ 20\AA $\times$ 2.44\AA. 
This corresponds to two layers of C atoms perpendicular to the
tube axis with twelve atoms per layer. 
We have studied three different H coverage, i.e., 
1, 2 and 24 H atoms 
per unit cell. We should point out that
due to the periodic boundary conditions along the tube axis, 1, 2
and 24 {\it rows} of H atoms are actually being simulated in the
calculations.
The diameter of the fully relaxed pure nanotube is found to
be 8.17 \AA.
The binding energy per H atom,  
\begin{equation}
E_b = 1/n[E_{tot}(tube) + E_{tot}(nH) - E_{tot}(tube+nH)]
\end{equation}
is calculated in terms of the total energy of the pure nanotube
$E_{tot}(tube)$, 
the total energy of the $n$ (=1, 2, 24) H atoms 
$E_{tot}(nH)$,
and the total energy of the nanotube with $n$H atoms, 
$E_{tot}(tube+nH)$.
Since $E_{tot}(nH)$ is calculated using the same 
supercell geometry as in the ($tube+nH$) system, 
the spurious 
adatom-adatom interactions along the tube axis are subtracted.
In Table I we list values of the binding energy per H 
atom at the interior and 
exterior of the nanotube for one H atom, 
a pair of H atoms on the same and adjacent layers perpendicular to the 
nanotube axis, as well as the fully hydrogenated case.
The positive value of $E_b$ in all cases indicates that the chemisorption  
is exothermic and hence energetically stable. 

First, we consider the H coverage of one atom per unit cell, with 
the H atom adsorbed 
either at the interior or exterior of the tube. 
By placing the H atom at various initial positions away from the nanotube wall,
we find that there is no energy barrier for H chemisorption 
at both inside and outside of the tube.
Fig. 1(a) shows the relaxed atomic structure with the H atom adsorbed
at the exterior to the tube. We label the nearest-neighbor C atom 
to the H by C$_{1}$, and the nearest-neighbor C atom to C$_{1}$ at
the same layer by C$_{2}$.
The calculated C$_{1}$-H bond length (1.11 \AA) is the same for the
H atom at either side of the nanotube wall, and it is
close to the corresponding value of 1.10 \AA~ in a CH$_4$ molecule.
However, there is a significant difference in the binding energy for H at
the interior (0.49 eV) and the exterior (1.77 eV) of the nanotube wall.
This is due to the fact that the chemisorption of H results in a transition 
from $sp^2$-like bonding in the pure nanotube to 
$sp^3$-like bonding in the presence of H; and  
the latter cannot be fully formed upon adsorption 
to the inner wall because of the unfavorable bonding angle \cite{tada}.
The Mulliken analysis of bond population \cite{mulliken,sanchez,segall} shows that 
0.33 electrons is transferred 
from the H atom to its nearest-neighbor C$_{1}$, in 
agreement with previous calculations \cite{ciraci} for the 
fully hydrogenated nanotubes.
The C$_1$- $s$ and $p$ orbitals acquire 0.15 and 0.18 electrons, respectively.
It is important to note that for the pure nanotube the
overlap population between the nearest-neighbor C atoms on the same and two adjacent layers is 0.87 and 2.22, respectively, indicating 
a stronger bonding in the latter case
\cite{sanchez}.
This large difference in overlap population
between the two types of nearest-neighbor C bonds
will in turn affect the propensity of H-induced  
breaking of the C-C bonds.
The chemisorbed H results in an increase of the 
C$_{1}$-C$_{2}$ bond 
length from 1.42 \AA~ (in the pure nanotube) to 1.53 \AA, with 
the latter value being typical for $sp^3$ C-C bonds.
As a result, the C$_{1}$-C$_{2}$ overlap population decreases 
from 0.87 to 0.63, indicating that the C-C bond is weakened.
In order to gain more insights into the change of the electronic structure
accompanying the H chemisorption process, we have calculated 
the angular momentum- and site- 
projected density of states (DOS) for some representative cases (Fig. 2).
For the pure nanotube, we have confirmed the metallic behavior of
the (6,6) armchair nanotube with 
a low density of states at the Fermi energy ($E_F$)
\cite{dresselhaus1}.   
From Fig. 2(a), we find that the H chemisorption 
induces the opening of a band gap ($\approx$ 5 eV) for
the C$_{1}$ atom, which is the
result of 
the large charge transfer from the H atom to C$_1$, filling 
up the C$_{1}$ $p$ valence band.
On the other hand, it is interesting to note that 
the C$_{2}$ atom  
exhibits a large narrow peak across $E_F$ which overlaps with the
bonding states of the H atom.
This peak, reminiscent of a Van Hove singularity, arises from 
the degeneracy of the two linear bands 
crossing $E_F$ in the pure nanotube, which is in turn lifted under the 
local distortion induced by H, thus forming two sub-bands whose
extremes fall at E$_F$.
Fig. 3(a) shows the valence charge density contour plot for a plane 
containing the C atoms on the same layer.
One can see the directional $sp^2$-like covalent 
bonding between the C atoms,
the $sp^3$-like bonding between C$_{1}$ and H, 
and the weakening of the C$_{1}$-C$_{2}$ $sp^2$ bond.

The result of the H-induced weakening of the 
C-C bond invites an interesting question:
Could the chemisorption of another H atom 
on the nearby C atom 
lead to the formation 
of two H-C bonds which will in turn further decrease or even
break the C$_{1}$-C$_{2}$ bond, resulting in
the unzipping of the entire nanotube? 
Thus, next we consider the case of a pair of H 
atoms chemisorbed on the C$_1$ and C$_2$ atoms in the exterior
and interior to the tube. We have examined
two different adsorption sites for the second H atom, either at
the same or the adjacent C layer. 
First, we present results with 
the interatomic axis between the H atoms 
parallel to the C$_{1}$-C$_{2}$ bond.  
The binding energy per H for this case 
is 0.37 eV (inside the tube) and 3.0 eV 
(outside the tube), respectively. 
As in the single H case, 
the pair of H 
atoms prefers to be chemisorbed at the exterior 
of the nanotube.  
Interestingly, 
the binding energy per H atom for the pair outside the nanotube 
has dramatically increased compared the corresponding value for the
single H case, while
the charge transfer from each H  
to its nearest-neighbor C remains about the same (0.30 electrons) as
in the one H case.
This charge transfer  
leads to an ionic-like repulsion between the H atoms 
and between 
C$_{1}$ and C$_{2}$.
Figure 1(b) shows the relaxed atomic structure for the case of two 
H atoms adsorbed at the exterior 
of the tube on the same layer.
The most remarkable 
feature in this figure is the {\it breaking} of the 
nearest-neighbor C$_{1}$-C$_{2}$ bond.
The equilibrium H-C, C$_{1}$-C$_{2}$, 
and H-H bond lengths are 1.09 \AA, 4.07 \AA, and 2.94 \AA, respectively.
Due to the periodic boundary conditions along the tube axis, the 
result implies scission of the nanotube wall along the tube axis,
i.e., the unzipping effect necessary for the experimentally observed 
coalescence phenomenon. 
The overlap population of C$_{2}$-H is
0.71, which is as large 
as that for the nearest-neighbor C-C bond at the same layer 
in the pure nanotube. 
Fig. 2(b) shows the 
DOS for one of the hydrogenated C atoms, C$_2$ 
(see also Fig. 1b) and its nearest-neighbor, C$_{3}$.
The DOS for C$_2$ is similar to that of C$_{1}$ 
in Fig. 2(a). Note the narrowing of the H $s$ band 
which exhibits bonding and antibonding states
separated by an energy gap of about 5 eV, 
indicates that the H atom is stabilized due to the 
hybridization between the $s$ states of H and the $p$ states 
from the nearby C$_2$ atom. 
From Fig. 3(b), one can see the $sp^3$-like bonding 
between the C and the H atoms, 
and the breaking of the C$_{1}$-C$_{2}$ bond.

We have also studied the case of two H atoms chemisorbed 
on the two nearest-neighbor C atoms located at adjacent
layers shown 
in Fig. 1(c). The binding energy per H atom 
is found to be 0.73 eV in the interior and 2.52 eV in the exterior 
of the tube, respectively.
In contrast to the previous case,  
we find that the nearest-neighbor C-C bond does not break. 
The equilibrium H-C, C$_{2}$-C$_{3}$, 
and H-H bond lengths are 1.10 \AA, 1.50 \AA, and 2.29 \AA, respectively.
There is a similar amount of charge transfer from the H atoms to each
nearest-neighbor C atoms with an overlap population of 0.74. 
However, in contrast to the previous case,
the C$_2$-C$_3$ overlap population of 1.59 is reduced considerably  
relative to that of 2.2 in the pure nanotube.  
By employing a larger supercell 
(four layers per cell), we find essentially the same results,
i.e., the C-C bond remains unbroken in the presence of H atoms.
This result can be understood from the fact that the 
nearest-neighbor C-C bonds
on adjacent layers are stronger 
than those on the same layer, as alluded earlier.

Interestingly, the C-C bond can not break when the two H atoms
chemisorb  inside the nanotube regardless of the adsorption
site. The C-C bond deforms nevertheless, with the bond length
increasing from 1.42 \AA~ (pure tube) to 1.52 \AA.
This again originates from the 
reduced propensity for $sp^3$ bonding at the interior of the 
nanotube \cite{tada}.
Overall, the internal surface of the nanotube is found to be 
less reactive 
than the external surface, in agreement with 
the general consensus \cite{haddon,cho}. 
Finally, our calculations for the fully endo- and exo-hydrogenated 
nanotube find 
no breaking of the C-C bonds and a lower binding energy per 
H atom of 0.30 eV and 0.84 eV, respectively.
We have also carried out calculations for a H$_2$ molecule approaching
the nanotube wall, both from the exterior and the interior,
with the H-H bond 
perpendicular to the nanotube wall. In agreement with previous 
calculations \cite{chan}, 
we find that no chemisorption occurs and the H$_2$ 
molecule remains intact.

In conclusion, we have studied the chemisorption properties 
of atomic H on a single-walled (6,6) armchair nanotube. 
We find that H atoms bind strongly to the nanotube through 
$sp^3$ bonding, with the binding energy at the exterior of
the tube being much greater than that at the interior.
For the first time, we predict that a pair of H atoms 
chemisorbed on two nearest-neighbor C atoms on the same layer outside
the nanotube catalyzes the
breaking of the C-C bond 
and leads 
to the unzipping of the nanotube. This H-induced
unzipping mechanism lends strong support 
to the recent 
experimental observations for the 
coalescence of single-walled nanotubes in the presence of 
atomic H.
On the other hand, 
the C-C bond does not break when H atoms are adsorbed 
at the interior of the tube.

We acknowledge support from 
the U. S. Army under Grant Nos. DAAD19-00-1-0049 and 
DAAG55-97-1-0247.

\begin{figure}[p]
\includegraphics[width=300pt]{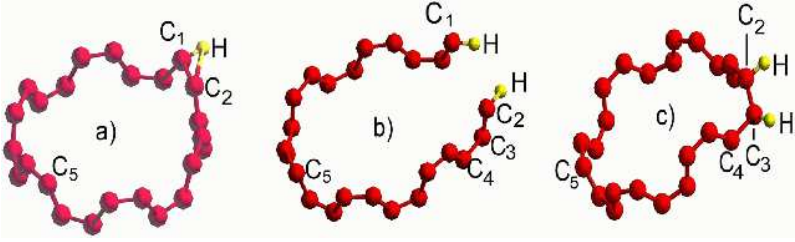}
\caption{Relaxed atomic structures for chemisorption of one H atom (a), 
and a pair of H atoms on the same layer (b) and on the
adjacent layers (c) of the (6,6) single-walled armchair nanotube.}
\end{figure}

\begin{figure}[b]
\includegraphics[width=300pt]{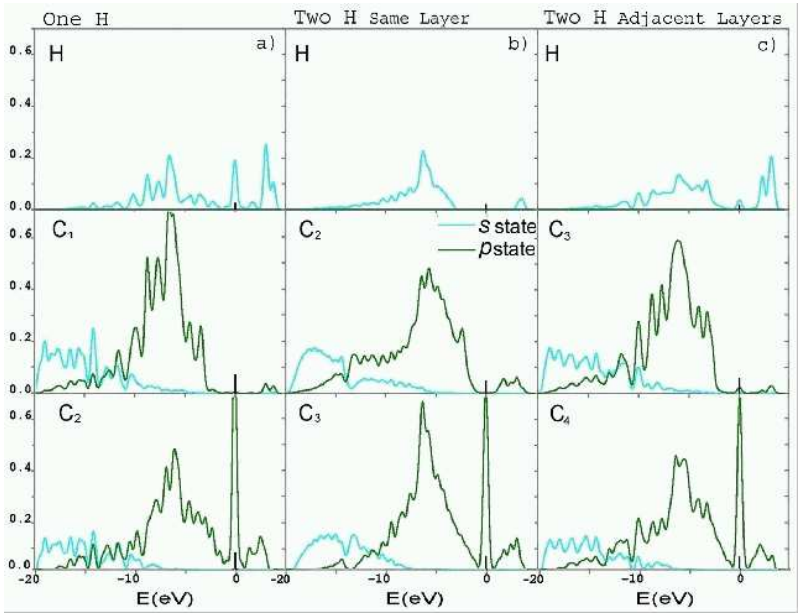}
\caption{The angular momentum and site projected DOS for the adsorption of 
one H atom (a), and a pair of H atoms on the same layer (b) 
and on the adjacent layers (c), exterior to the nanotube. The 
labeling of the various atoms is the same as that used in 
Figs. 1 (a), (b) and (c), respectively.}
\label{2}
\end{figure}

\begin{figure}
\includegraphics[width=300pt]{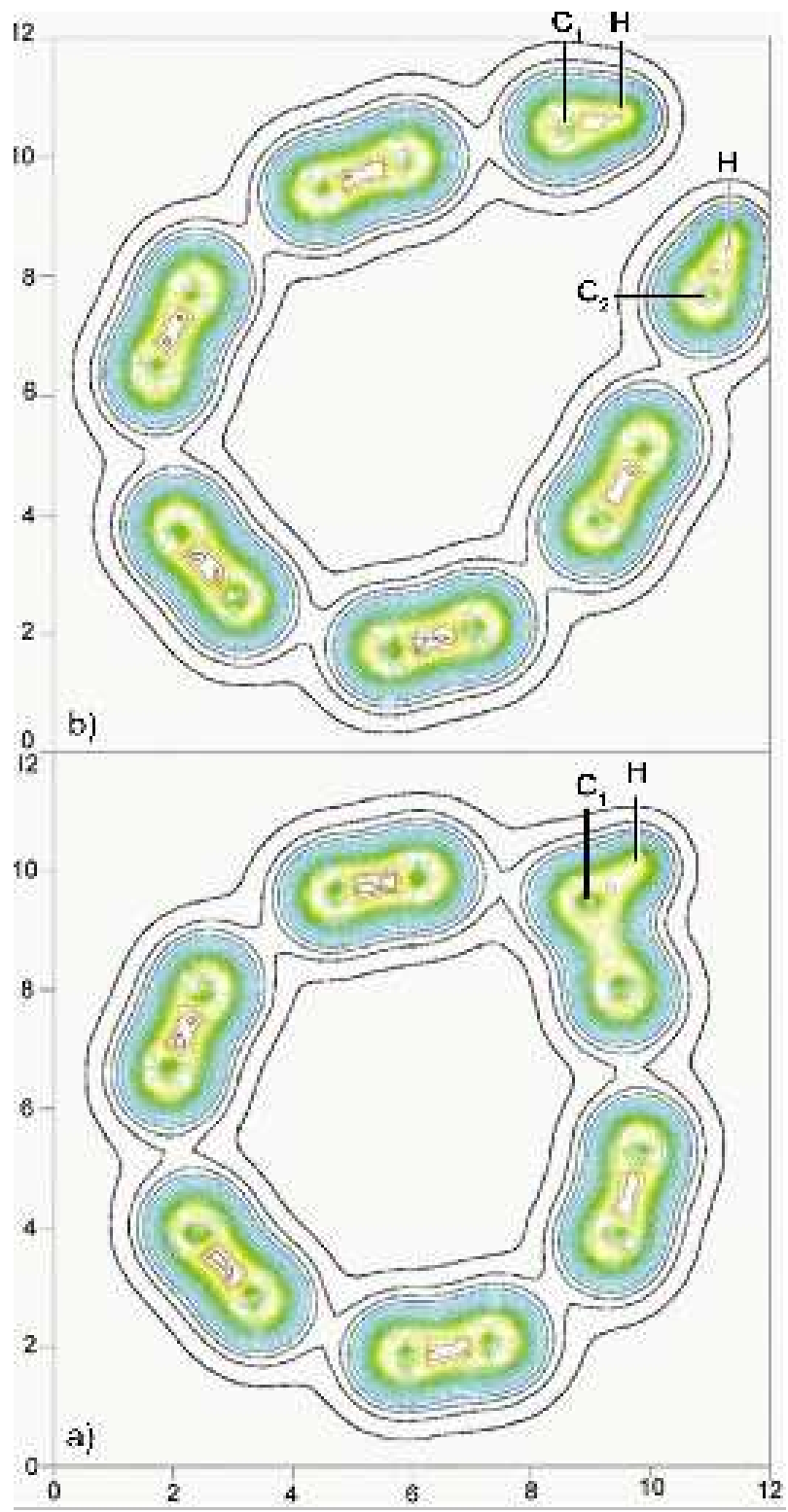}
\caption{The valence charge density contour plot
on a plane perpendicular to the nanotube axis passing through the C 
atoms, for adsorption of 
one H atom (a), and two H atoms on the same layer
(b).}
\label{3}
\end{figure}

\begin{table}[p]
\caption{Binding energy (eV) per H atom for chemisorption exterior 
and interior to the nanotube for one H atom, a pair of H atoms 
on the same and the adjacent layers, as well as the fully hydrogenated case.}
\begin{ruledtabular}
\begin{tabular*}{\columnwidth}{@{\extracolsep{\fill}}ccccc}
& 1 H  & 2 H same layer  & 
2 H adjacent layers & 24 H \\ \hline
interior & 0.49  & 0.37  & 0.73 &  0.30 \\
exterior & 1.77  & 3.00  & 2.52 & 0.84 \\
\end{tabular*}
\end{ruledtabular}
\end{table}
\end{document}